# Stutter Diagnosis and Therapy System

## Based on Deep Learning

Dr. Mrs. Gresha Bhatia[1], Binoy Saha[2], Mansi Khamkar[3], Ashish Chandwani[4], Reshma Khot[5]

Deputy HOD, CMPN department, Vivekanand Education Society's Institute of Technology (V.E.S.I.T), Chembur, Mumbai, India [1]

Student of Computer Enginnering, VESIT, India [2 3 4 5]

**ABSTRACT** — Stuttering, also called stammering, is a communication disorder which breaks the continuity of the speech. This program of work is an attempt to develop automatic recognition procedures to assess stuttered dysfluencies and use these assessments to filter out speech therapies for an individual. Stuttering may be in the form of repetitions, prolongations or abnormal stoppages of sounds and syllables. Our system aims to help stutterers by diagnosing the severity and type of stutter and also by suggesting appropriate therapies for practice by learning the correlation between stutter descriptors and effectiveness of speech therapies on them. This paper focuses on implementation of stutter diagnosis agent using Gated Recurrent CNN on MFCC audio features and therapy recommendation agent using SVM. It also presents the results obtained and various key findings of the system developed.

## I. Introduction

Stammering or stuttering is a disorder of speech which badly affects the speech fluency of the person. There are stoppages and disruptions pauses which interrupt or disturbs the fluency of speech. Stuttering may be in the form of repetitions of sounds, syllables or words - like saying mo-mo-mobile. There may also be prolonged sounds - like saying mmmmmmmobile. Sometimes no sound is heard due to silent blocking. Stuttering interferes with work and social life of an individual and often brings tremendous emotional suffering. According to research, more than 70 million people in the world stutter. Stuttering therapy includes various treatment methods that are used to reduce stuttering to some degree in an individual. Generally in stuttering detection process speech is recorded and disfluencies like repetitions, prolongation, interjection are identified. Then the disfluencies that occur are counted, according to that severity of stuttering is determined. Speech therapists use different approaches such as Lidcombe approach, stuttering modification, fluency shaping, Modifying Phonation Intervals (MPI), psychological therapies, and auditory feedback devices to treat stuttering and often combine several methods to meet individual needs. While it is difficult to eliminate stuttering, speech therapy helps the majority of children and adults to palliate its severity. According to the survey 84% people experienced improvement in fluency of speech. Also few adults (73 out of the surveyed people) have used assistive speech fluency devices, but they did not work well for more than 52% of them. [1][2]

### A. Problems

Private speech therapy is costly and not affordable for most families living in the poorer districts. The lack of education and training about the disorder of stuttering by professional adults, including speech therapists, doctors and educators, has tragic results. The speech therapy needs to be intense for two/three months and there needs to be a maintenance phase that is extended over a period of one year minimum so stutters have to visit the therapy centre each time. There is no way to judge the effectiveness of the homework given to stutters but it is very important because most of the people stutter in real world situations. Also the judgements made by one Speech Lab Pathologist (SLP) may differ from the judgements made by another SLP. The speech therapies are given randomly by the SLPs as there is no proper way to customize them by assessing the effectiveness of the therapies. [2][3][4]

**B. Need for technology**

Currently most SLPs don't use much technology, all the therapies are performed under therapist's guidance only. The SLPs have to manually listen to all recordings multiple times to jot down and count the words in which the patient has stuttered. Also there is no way to monitor stutterer's performance during practice or in public environment. Doing quantitative analysis of the patient's speech and learning which therapies are best suited for the patient according to his performance would pacify the treatment process. So technology here can prove helpful in automating several tasks to get better results. [1][2][3][4]

**C. Our focus**

This project intends to deliver an affordable personalized stuttering therapy to people who stutter. The main objective of this project is to improve person's speech fluency by accurately diagnosing stutter and then suggesting appropriate training exercises for practice. The system will continuously monitor user's performance and will recommend new tests accordingly to make sure that the tests are effective. Thus, the main goal of our work is to :
1. Provide quantitative analysis of the type and severity of stuttered disfluencies in speech
2. Learn correlation between stutter descriptors and various therapies available

## II. Previous Work

There has been research going on in this field to automate the process of stutter identification. But the idea of mapping stuttered dysfluency measurements with the different speech therapies to determine their effectiveness remains untouched. To summarize, existing research literature tells us that we can use ANNs, HMMs and SVMs to classify stuttered speech and non-stuttered speech with considerable accuracy (greater than 90%) as shown in the table below. [5][6][7][8]

| Year | Database | Features | Classifiers | Results(%) |
|---|---|---|---|---|
| 2009 | 10 samples from UCLASS | MFCC | K-NN, LDA | 90.91% |
| 2009 | 10 samples from UCLASS | LPCC | K-NN, LDA | 89.77% |
| 2010 | 2 speakers | MFCC | HMMs | 80% |
| 2010 | 121 speakers | Time Domain, Spectral domain | Batesian detector, HMM,LDA | 63% |
| 2012 | 39 samples from UCLASS | MFCC | k-NN,LDA | 92.55% |
| 2012 | 39 samples from UCLASS | LPC,LPCC,WLPCC | k-NN | 92.16% 96.47% 97.45% |
| 2012 | 39 samples from UCLASS | LPC,LPCC,WLPCC | LDA | 94.90% 97.06% 98.04% |
| 2012 | 53 speakers | Spectral measures(FFT 512) | ANNs | 84% |
| 2013 | 39 samples from UCLASS | LPCC,WLPCC,MFCC | SVM | 95% |
| 2014 | 16 samples from UCLASS | MFCC | SVM | 98.00% |
| 2014 | 10 speakers | volume, zero crossing rate, spectral entropy, high-order derivatives, VH curve, and VE curve and end-point detection according (EPD) | DTW | 83% |
| 2015 | 50 samples from UCLASS | MFCC | GMM | 96.43% |

*Table 1. Previous work on stutter detection*

## III. Our Approach

The basic blocks and flow of the system is shown in figure 1. The system consists of 2 main modules - stutter assessment and therapy suggestion. Both are developed using python and its libraries. Our system recognizes 2 types of stutter in speech recordings - prolongation and repetition and calculates its severity index.

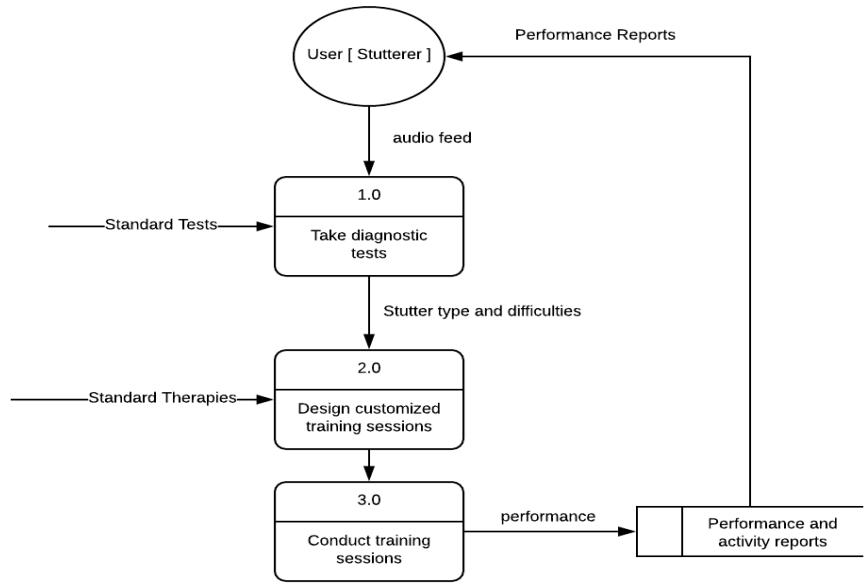

*Fig 1 : Basic blocks and flow of Stutter Therapy System*

**A. Stutter Assessment**
This module provides quantitative analysis of the stutter present in the recorded speech. The procedure followed is as explained below

a. Dataset
The dataset used by us to train our models to identify stutter is the University College London's Archive of Stuttered Speech (UCLASS). It contains recordings of actual stutterers of all ages and varying severity levels. We chose their 2nd release prepared in 2008 due to its more recent nature and good quality audios. There are 3 categories of recordings provided by this release - monologue, reading and conversation out of which we used recordings under monologue and reading. The monologue folder consists of 82 recordings (6 female and 76 male) and the reading folder consists of 108 recordings (15 female and 93 male). We assume that this imbalance in female and male data samples is purposefully made due to the fact that stuttering is more common among boys than girls, the male-to-female ratio is large, about 4 to 1 or greater. The audio files were in 2 formats - .mp3 and .wav. We preferred .wav over .mp3 as .wav files are lossless and uncompressed thus maintaining the original quality of the recordings.

b. Data Labelling
We listened to all the recordings from the downloaded dataset and manually trimmed the best non-stuttered and stuttered speech parts out of them. The stuttered trimmed audios contain 2 sets - prolongation and repetition with varying severity levels. There is one common set of non-suttered trimmed audios which is created from the best fluent speech parts from the dataset and some recordings downloaded randomly from various sources available on the internet. This set also contains few samples of only background noise. These trimmed audios are then labelled as stutter or non-stutter separately for prolongation and repetition. [5][9][10][1]

c. Feature Extraction
The trimmed audios are further cut into fixed sized segments of 1 second each and each segment is given the label of its respective trimmed audio. Segmentation is done using python's PyDub library. We then extracted MFCC features of 13 coefficients for each segment and obtained feature arrays of size (13,44) using python's Librosa library. These MFCC feature arrays are our actual data points for training. We preferred MFCC features over others as the Mel scale relates perceived frequency (pitch) of a pure tone to its actual measured frequency which makes our features match more closely with what humans hear. [12][13][14][15][16]

d. Training

We trained 2 separate models - one for identifying prolongation and other for identifying repetition as both showed different patterns in the audio signals. We trained model for prolongation on only 1st and 13th MFCC coefficients whereas the model for repetition was trained on all the 13 MFCC coefficients. Both models were built using Gated Recurrent Convolutional Neural Network (GRCNN) architecture. They gave us the best validation accuracy of about 90% which we further improved to ~92% for repetition and ~95% for prolongation after tweaking various hyper-parameters. The other algorithms which we tried before GRCNN are - SVM with both linear and non-linear kernels, HMM, multilayer neural network, CNN and RNN listed in increasing order of their accuracy. But GRCNN out performed them all as it combines the advantages of both CNN and RNN. CNN extracted repetition and prolongation patterns in 2D MFCC feature array and the Gated Recurrent Units (GRUs) predicted stutter by analysing previous sequences. GRUs are improved version of standard recurrent neural network which solve the problem of vanishing gradient using memory and gates like update and reset. The model is trained with hyper-parameters such as relu activation function, sigmoid activation function at the end, binary cross entropy loss and batch size 64. We used Keras with tensorflow backend to develop and train these models. The architecture for both models are shown below. [17][18]

| Layer (type) | Output Shape | Param # |
| --- | --- | --- |
| conv2d_17 (Conv2D) | (None, 2, 20, 32) | 192 |
| activation_25 (Activation) | (None, 2, 20, 32) | 0 |
| conv2d_18 (Conv2D) | (None, 2, 8, 32) | 5152 |
| activation_26 (Activation) | (None, 2, 8, 32) | 0 |
| reshape_9 (Reshape) | (None, 16, 32) | 0 |
| gru1 (GRU) | (None, 16, 32) | 6240 |
| gru2 (GRU) | (None, 32) | 6240 |
| dropout_9 (Dropout) | (None, 32) | 0 |
| dense_9 (Dense) | (None, 1) | 33 |
| activation_27 (Activation) | (None, 1) | 0 |

Total params: 17,857
Trainable params: 17,857
Non-trainable params: 0

None

| Layer (type) | Output Shape | Param # |
| --- | --- | --- |
| conv2d_120 (Conv2D) | (None, 3, 37, 32) | 288 |
| activation_151 (Activation) | (None, 3, 37, 32) | 0 |
| conv2d_121 (Conv2D) | (None, 3, 30, 32) | 8224 |
| activation_152 (Activation) | (None, 3, 30, 32) | 0 |
| conv2d_122 (Conv2D) | (None, 3, 23, 48) | 12336 |
| activation_153 (Activation) | (None, 3, 23, 48) | 0 |
| conv2d_123 (Conv2D) | (None, 3, 16, 48) | 18480 |
| activation_154 (Activation) | (None, 3, 16, 48) | 0 |
| conv2d_124 (Conv2D) | (None, 3, 9, 64) | 24640 |
| activation_155 (Activation) | (None, 3, 9, 64) | 0 |
| reshape_36 (Reshape) | (None, 27, 64) | 0 |
| gru1 (GRU) | (None, 27, 32) | 9312 |
| gru2 (GRU) | (None, 32) | 6240 |
| dropout_32 (Dropout) | (None, 32) | 0 |
| dense_32 (Dense) | (None, 1) | 33 |
| activation_156 (Activation) | (None, 1) | 0 |

Total params: 79,553
Trainable params: 79,553
Non-trainable params: 0

None

*Fig 2 : Left - Architecture of GRCNN model to detect prolongation,*
*Right - Architecture of GRCNN model to detect repetition*

**B. Therapy Suggestion**

This module recommends an appropriate set of therapies to the patient according to his performance over time and speech fluency improvement. The details about this module are given below :

a. Dataset

As this idea of finding correlation between stutter descriptors and speech therapies is unexplored, there is no such dataset available yet. The dataset will be correctly generated when the performance and improvement of the stutterers will be recorded for each therapy. A model can then be trained on this generated dataset to learn which set of therapies should be suggested to the patient to pacify his improvement in speech fluency by observing his previous performance results. Initially, instead of providing random therapies to the patient, we manually developed a small dataset to train the model based on what research scholars had written in their articles and our intuition. The dataset consists of parameters like prolongation index, repetition index and speech fluency improvement index over time. The labels are the names of various speech therapies available, each therapy further divided into 3 levels - easy, medium and hard. This dataset indicates that if the stutter severity index is low with high improvement, then difficult therapies should be suggested and if stutter severity index is high with low improvement, then easy therapies should be suggested. A part of the initial dataset is shown in figure 3, where therapy names as labels are one hot encoded and the values for prolongation, repetition and improvement indicate -   1 : < 25%,   2 : 25% - 50%,   3 : 50% - 75%,   4 : > 75%

For example, as shown in figure 3, there exists a pattern such that if prolongation is very low, then therapy 1 should not be suggested and if repetition index is high with low improvement, only then therapy 2 should be suggested.

| prolongation | repetition | Improvement | Therapy 1 | Therapy 2 |
|---|---|---|---|---|
| 1 | 1 | 1 | 0 | 0 |
| 1 | 1 | 2 | 0 | 0 |
| 1 | 1 | 3 | 0 | 0 |
| 1 | 1 | 4 | 0 | 0 |
| 1 | 2 | 1 | 0 | 0 |
| 1 | 2 | 2 | 0 | 0 |
| 1 | 2 | 3 | 0 | 0 |
| 1 | 2 | 4 | 0 | 0 |
| 1 | 3 | 1 | 0 | 1 |
| 1 | 3 | 2 | 0 | 0 |
| 1 | 3 | 3 | 0 | 0 |
| 1 | 3 | 4 | 0 | 0 |
| 1 | 4 | 1 | 0 | 1 |
| 1 | 4 | 2 | 0 | 1 |
| 1 | 4 | 3 | 0 | 0 |
| 1 | 4 | 4 | 0 | 0 |

Fig 3 : A part of the initial dataset to train the therapy suggestion model

b. Training

Once the initial dataset was ready, we trained an SVM model with polynomial kernel on it using scikit-learn, achieving an accuracy of about 94%. This model can also be trained on the real dataset that will be further generated as the performance of the patients is monitored.

## IV. Evaluation

**A. Calculate stutter severity index**

$$\left( \frac{\text{Number of segments classifies as stutter}}{\text{Total number of audio segments}} \right) \times 100$$

**B. Calculate improvement in speech fluency**

$$\text{Ceil}\left( \frac{\text{Avg}\{(I_p - C_p) + (I_r - C_r)\}}{25} \right)$$

Where -
$I_p$ : Initial prolongation severity index obtained at the time of diagnosis
$C_p$ : Current prolongation severity index
$I_r$ : Initial repetition severity index obtained at the time of diagnosis

Cr : Current repetition severity index

The improvement values are squashed between 1-4 which determine the levels of improvement in percentage as shown below :

1 : < 25%,   2 : 25% - 50%,   3 : 50% - 75%,   4 : > 75%

## V. Key Findings

1. MFCC features give best results with deep learning models.

We found that for us deep learning models outperformed other models such as SVM and HMM. GRCNN gave us best accuracy as it combines the advantages of both CNN and RNN. The 2 GRCNN models trained separately for recognising prolongation and repetition in speech audio achieved validation accuracy of 95% and 92% respectively by further tweaking certain hyperparameters. The following table shows all the models that we have tried and their results.

| | Model | Findings | Validation Accuracy |
|---|---|---|---|
| Other Models | SVM with linear kernel | Dataset was not linearly separable | ~ 54% |
| | SVM with non-linear kernel | Incremental learning not supported | ~ 63% |
| | HMM | Difficult to determine states | ~ 65% |
| DL Models | Multi-layer NN | Could not find patterns in speech | ~ 67% |
| | Convolutional NN | Filters could capture some patterns | ~ 80% |
| | Recurrent NN | Processed data in time series but was unstable | ~ 78% |
| | Recurrent CNN | Benefits of both CNN and RNN boosted accuracy | ~ 90% |

*Fig 4 : Various models we trained with their validation accuracies*

2. Single model could not identify all types of stutter.

We first tried to detect prolongation and repetition using a single GRCNN model. Although it gave us validation accuracy of 84%, it gave unsatisfactory results on testing audio files. Realizing that each type of stutter formed different patterns in audio signal, we trained 2 separate GRCNN models for detecting prolongation and repetition.This boosted the accuracy to 92% for repetition and 95% for prolongation.

3. Skew dataset with large number of samples of class non-stutter and lesser number of samples of class stutter improved accuracy.

We realized that in the real world, the cases of non-stutter are much more than the cases of stutter, So the dataset should be binary asymmetric. Thus, we brought the ratio of stuttered speech samples to non-stuttered speech samples from 50:50 to 25:75 percent. This avoided the problem of predicting false positives.

4. Audio segments of length 1 second gave best results.

To train the GRCNN model, we first trimmed the best stuttered and non-stuttered speech parts out of the recordings in dataset. Then we segmented these trimmed audios into equal sizes and labelled each segment as stutter or non-stutter. Different segment sizes were tried such as 0.2 sec, 0.5 sec, 0.7 sec, 1 sec, 1.5 sec, 2 sec among which the size of 1 sec worked best for us.

5. The trained model is also considering the underlying voice quality along with the presence or absence of stutter.

While testing our models on random speech recordings by actual stutterers as well as artificial stutterers, we observed that the models always predicted low stutter severity index on artificial stuttered speech than that predicted on the recordings by actual stutterers. This indicates that the models have also learnt to consider the underlying voice quality of speaker. This is because the voice quality of a natural stutterer is consistently bad (shaky voice) unlike an artificial stutterer.

6. MFCC coefficients 1 and 13 clearly showed a pattern for prolongation.
For analysing the MFCC feature arrays of prolonged speech, we picked a few audios and plotted graphs of non-stuttered MFCC features vs prolongation MFCC features. We noticed that the 1st and 13th MFCC coefficients showed clear patterns in the graph as displayed below. Thus, we trained our model for prolongation on only the 1st and 3rd MFCC coefficients which reduced each feature array to size (2,44). This further improved the accuracy of our model for detecting prolongation.

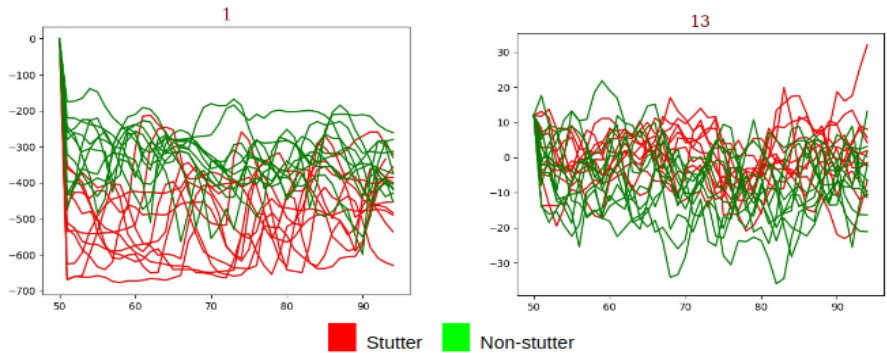

*Fig 5 : Patterns in MFCC features for prolongation vs non-stuttered speech samples*

7. No pattern was found in MFCC coefficients for repetition.
Same approach was used to discover patterns in MFCC feature arrays of speech containing repetition, but no clear pattern was found. Hence, we trained our model to detect repetition on all 13 MFCC coefficients, which left the size of MFCC arrays (13,44) unchanged.

## VI. Results

The 2 GRCNN models trained separately for recognising prolongation and repetition in speech audio achieved validation accuracy of 95% and 92% respectively. The accuracy of these models was increased by introducing an imbalance in the dataset (with large number of samples of class non-stutter and lesser number of samples of class stutter), fixing the length of audio segments to 1 second, selecting only those MFCC coefficients which showed clear patterns and tweaking the hyperparameters of the models.  Also, 94% validation accuracy is achieved by the SVM model trained to recommend best suited therapies.

## VI. Conclusion

In this paper we have discussed the approach we used to automatically identify stuttered dysfluencies and learn the effectiveness of the various speech therapies. Perhaps a great deal of our energy and time was spent in listening, trimming and labelling the audio files due to the lack of well-structured dataset available for stuttered speech. The biggest impediment to our models was that the background noise messed up the speech in the audio signals which resulted in high number of false positives. We tried various classification algorithms, visualized MFCC features to look for speech patterns, augmented the models and data samples to reach the current accuracy level. Hence, these trained models are now integrated in a full-fledged mobile application  which can be used by stutterers for speech therapy.

## VIII. Future Scope

Currently, we have our models trained to identify prolongation and repetition as well as a model to suggest appropriate therapies. Following are some additions which we wish to develop in future :
- To identify other types of stuttered dysfluencies such as blockages, insertions etc.
- To include all speech therapies available and learn their correlation with stutter descriptors.
- To completely remove or avoid background noise so that accurate results are predicted.